\documentclass[10pt,aps,prd,twocolumn,noshowpacs,superscriptaddress]{revtex4-1}

\usepackage{mathrsfs, bm,color,amsmath,amssymb, stmaryrd, amsfonts,latexsym,graphicx, psfrag}
\usepackage{bbold}	
\usepackage[colorlinks=true,citecolor=green,linkcolor=red,urlcolor=blue]{hyperref}

\newcommand{\nn}{\nonumber}
\newcommand{\be}{\begin{equation}}
\newcommand{\ee}{\end{equation}}
\newcommand{\ba}{\begin{eqnarray}}
\newcommand{\ea}{\end{eqnarray}}

\newcommand{\remove}[1]{}

\begin{document}

\title{Orbital embedding and topology of one-dimensional two-band insulators}

\author{Jean-No\"el \surname{Fuchs}}
\email{fuchs@lptmc.jussieu.fr}
\affiliation{Sorbonne Universit\'e, CNRS, Laboratoire de Physique Th\'eorique de la Mati\`ere Condens\'ee, LPTMC, 75005 Paris, France}
\author{Fr\'ed\'eric \surname{Pi\'echon}}
\email{piechon@lps.u-psud.fr}
\affiliation{Universit\'e Paris-Saclay, CNRS, Laboratoire de Physique des Solides, 91405, Orsay, France}

\date{\today}
\begin{abstract}
The topological invariants of band insulators are usually assumed to depend only on the connectivity between orbitals and not on their intra-cell position (orbital embedding), which is a separate piece of information in the tight-binding description. For example, in two dimensions, the orbital embedding is known to change the Berry curvature but not the Chern number. Here, we consider one-dimensional inversion-symmetric insulators classified by a $\mathbb{Z}_2$ topological invariant $\vartheta=0$ or $\pi$, related to the Zak phase, and show that $\vartheta$ crucially depends on orbital embedding. We study three two-band models with bond, site or mixed inversion: the Su-Schrieffer-Heeger model (SSH), the charge density wave model (CDW) and the Shockley model. The SSH (resp. CDW) model is found to have a unique phase with $\vartheta=0$ (resp. $\pi$). However, the Shockley model features a topological phase transition between $\vartheta=0$ and $\pi$. The key difference is whether the two orbitals per unit cell are at the same or different positions. 
\end{abstract}


\maketitle


\section{Introduction \label{sec:intro}}

The canonical example of a topological insulator is the two-dimensional Chern insulator. Its occupied bands have a non-zero total Chern number, which leads to a quantized Hall effect. A simple example is Haldane's two-band model on the honeycomb lattice~\cite{Haldane1988}. Such a tight-binding model is defined by a state space spanned by orbitals, by a Hamiltonian giving the connectivity between orbitals and by a position operator giving their spatial embedding. Geometrical quantities such as the Berry curvature depend on the orbital embedding~\cite{Fuchs2010,Fruchart2014,Lim2015,Simon2020}. In contrast, the Chern number, a topological quantity which is the integral of the Berry curvature over the whole Brillouin zone (BZ), does not. It is therefore commonly assumed that topological invariants are generically independent of orbital embedding and only depend on the connectivity between orbitals. In other words, topological quantities depend on the Hamiltonian topology and geometrical quantities depend in addition on the position operator (see e.g.~\cite{Simon2020}).

In the ten-fold classification of topological insulators and superconductors~\cite{Schnyder2008,Kitaev2009,Chiu2016}, the starting point is a $k$-periodic Bloch Hamiltonian, where $k$ spans the BZ. It is well-known that there are two main conventions for the Bloch Hamiltonian, essentially depending on a choice in the Fourier transform~\cite{Bena2009,Fuchs2010,Fruchart2014,Lim2015,Vanderbilt2018}. In one convention, the intra-cell position of orbitals is irrelevant and the resulting Bloch Hamiltonian is always periodic with the BZ. However, this \textit{periodic} Bloch Hamiltonian is not unique and depends on a choice of unit cell. In the other convention, the intra-cell positions are taken into account, the Bloch Hamiltonian is unique and almost never periodic (except if there is a single site per unit cell hosting all the orbitals). We call it the \textit{canonical} Bloch Hamiltonian. The different Bloch Hamiltonians are related by a $k$-dependent gauge transformation involving the distances between orbitals within the unit cell. See Appendix~\ref{sec:now} for details and~\cite{Cayssol2021} for a recent review. Cases where it is crucial to retain the non-periodicity of the Bloch Hamiltonian therefore may escape this ten-fold classification.

In the present article, we consider one-dimensional band insulators and show that their topological characterization crucially depends on the orbital embedding and can not be obtained from the knowledge of the tight-binding Hamiltonian alone. As an emblematic example, we show that the SSH model~\cite{SSH} can either have a single trivial phase or two phases separated by a topological transition, depending on the orbital embedding. The first situation corresponds to the original SSH model in which the two orbitals are located on different sites (we call it SSH$_{1/2}$, where $d=1/2$ refers to the distance between the two sites, or simply SSH)~\cite{SSH}. The second situation corresponds to a modified SSH model in which the two orbitals would be on the same site (we call it SSH$_0$, as $d=0$). We show that SSH$_{1/2}$ and SSH$_0$ are physically different and are not topologically equivalent.

\begin{figure}[t]
\begin{center}
\includegraphics[width=\linewidth]{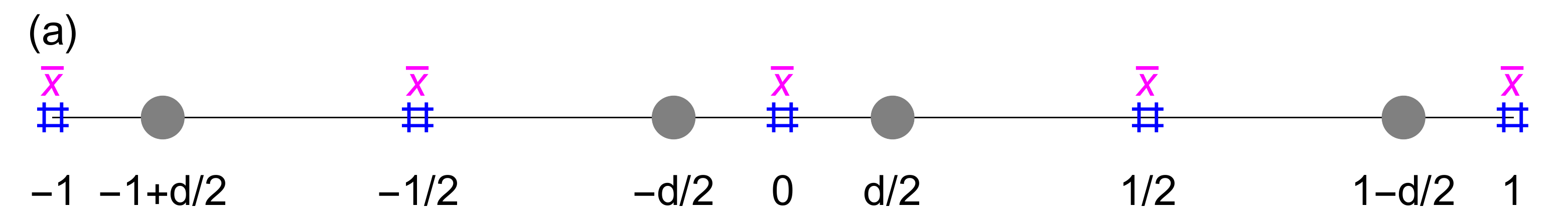} 
\includegraphics[width=\linewidth]{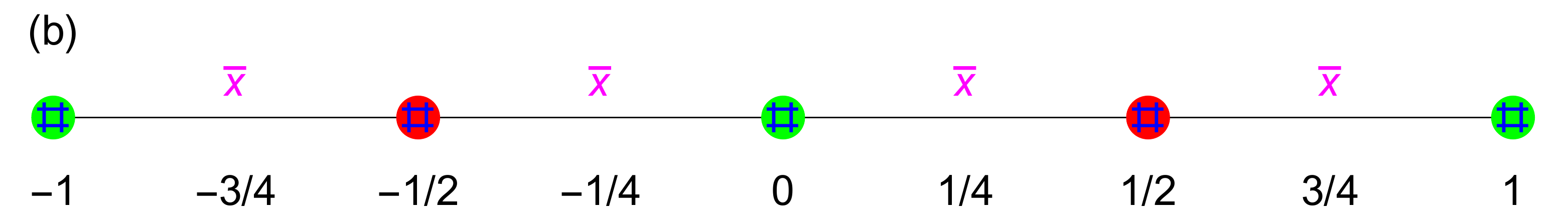} 
\includegraphics[width=\linewidth]{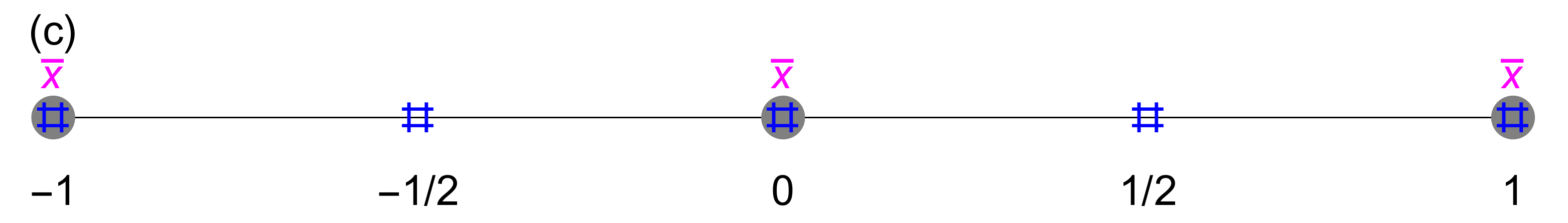} 
\caption{\label{fig:inversion} 1D crystals with (a) bond inversion, (b) site inversion, and (c) mixed inversion. Identical sites are shown as gray dots and non-equivalent ones as red and green dots. Inversion centers are indicated by blue \# and the average site positions in the unit cell in magenta $\bar{x}$.}
\end{center}
\end{figure}

Here, we restrict ourselves to inversion-symmetric insulators, which are characterized by a $\mathbb{Z}_2$ topological invariant $\vartheta=0$ or $\pi$~\cite{Hughes2011,ChenLee2011}, closely related to the Zak phase~\cite{Zak1989, Zak1989b}. Inversion-symmetric crystals have two inversion centers per unit cell. Depending on their position with respect to the sites, three situations arise: bond, site or mixed inversion (see Fig.~\ref{fig:inversion}). As representative two-band examples, we consider the SSH$_{1/2}$ model (bond inversion)~\cite{SSH}, the charge density wave (CDW) model (site inversion) and the Shockley model of coupled $s$ and $p$ bands (mixed inversion)~\cite{Shockley,Vanderbilt1993}. The SSH$_0$ model has mixed inversion symmetry and will also be briefly discussed. 

\medskip

The article is organized as follows. In Sec.~\ref{sec:zak}, we define a modified Zak phase that is position-origin independent, in contrast to the standard Zak phase. Then, in Sec.~\ref{sec:isi}, we review the $\mathbb{Z}_2$ classification of inversion-symmetric insulators. Next, we study the SSH$_{1/2}$ model in Sec.~\ref{sec:ssh}, the CDW model in Sec.~\ref{sec:cdw} and the Shockley model in Sec.~\ref{sec:shockley}. Eventually, we give a conclusion and perspectives (see Sec.~\ref{sec:ccl}). In Appendix~\ref{sec:now}, we review two conventions for writing a Bloch Hamiltonian when there are several sites per unit cell and explain why a bulk winding number can be defined for the SSH$_{0}$ model, but not for the SSH$_{1/2}$ model despite its chiral symmetry. In Appendix~\ref{sec:ep}, we relate the present study of the Zak phase (that focuses on electronic properties) to that of the electric polarization (that also involves ions). In Appendix~\ref{sec:rm}, we study the Rice-Mele model~\cite{RiceMele1982}, which is a generalization of the SSH$_{1/2}$ and CDW models that breaks inversion symmetry and compute its Zak phase analytically. 
%
%
%


\section{Zak phase\label{sec:zak}}
The Zak phase~\cite{Zak1989} is a peculiar Berry phase (Zak calls it non-cyclic~\cite{Zak1989b} and Resta an open-path Berry phase~\cite{Resta2000}) defined for a given band $n$ along a non-contractible loop of the first BZ. It reads
\begin{eqnarray}
Z_n = \int_{-\pi}^\pi \hspace{-0.2cm} dk \langle u_n |i \partial_k u_n\rangle + \text{arg} \langle u_n(-\pi)|e^{i 2\pi x} |u_n(\pi)\rangle ,
\label{eq:zak}
\end{eqnarray}
where $x$ is the position operator (we have taken a unit Bravais lattice spacing) and $|u_n(k)\rangle$ are the cell-periodic Bloch states, i.e. the eigenvectors of the canonical Bloch Hamiltonian $H(k)$, see Appendix~\ref{sec:now}. This expression is valid for periodic boundary conditions (PBC), in the thermodynamic limit and in any gauge, thanks to the arg$\langle...\rangle$ term in (\ref{eq:zak})~\cite{footnotegauge}. The key property of the Zak phase is that it is proportional to the Wannier (or band) center $\langle x_n \rangle$ of the band: $Z_n=2\pi \langle x_n \rangle$~\cite{Zak1989, Zak1989b}. 

Because of its relation to the position operator, the Zak phase depends continuously on the choice of position origin: when $x\to x+x_0$, where $x_0$ is an arbitrary real number, $Z_n\to Z_n+2\pi x_0$. Invariance under translation by a Bravais lattice spacing $x_0=1$ shows that $Z_n$ is defined mod $2\pi$. 

In order to remove this dependence on the position origin, we define the ``modified Zak phase'' by 
\begin{eqnarray}
\bar{Z}_n=Z_n-2\pi \bar{x}=2\pi (\langle x_n \rangle - \bar{x}), 
\label{eq:zakbar}
\end{eqnarray}
where $\bar{x}$ is the mean position of sites in a unit cell. Physically, $\langle x_n \rangle - \bar{x}$ measures the position of the Wannier center (i.e. of the localized electrons filling the $n^\text{th}$ band) with respect to the average position of sites.

A subtle issue is that, even if $Z_n$ is mod $2\pi$, $\bar{Z}_n$ is not necessarily defined mod $2\pi$. Let us call $Z_q$ the ``quantum of Zak phase'', such that $\bar{Z}_n$ matters mod $Z_q$. The quantum of Zak phase reflects the behavior of $\bar{Z}_n$ under space translation, which involves both $Z_n$ and $\bar{x}$. If there is a single site per unit cell then $\bar{x}=x_0$ mod $1$, where $x_0$ depends on the choice of origin, and therefore $Z_q=2\pi$. But if there are two sites per unit cell (as in the SSH$_{1/2}$ model), then $\bar{x}=x_0$ mod $1/2$ and then $Z_q=\pi$ (see Sec.~\ref{sec:ssh} for a detailed proof).

The modified Zak phase $\bar{Z}_n$ and the quantum of Zak phase $Z_q$ are closely related to the bulk electric polarization $P$ of crystals and the corresponding quantum of polarization $P_q$~\cite{King-Smith1993,Vanderbilt1993,Resta2007,Vanderbilt2018}, see also Appendix~\ref{sec:ep}. There is a subtle difference however. The modified Zak phase does not depend on a model for ions in contrast to $P$. It is purely an electronic quantity that depends on the tight-binding Hamiltonian (hopping amplitudes, connectivity of orbitals) and on the position operator (orbital embedding, i.e. position of sites in the tight-binding model)~\cite{footnoteposition}, but not on the charge or position of ions.

The two key differences between $\bar{Z}_n$ and $Z_n$ are that the former is independent of the position origin and defined mod $Z_q$, whereas the latter depends continuously on the position origin and is defined mod $2\pi$. 
Throughout the article, unless otherwise specified, we choose the position origin such that $\bar{x}=0$ mod $x_q$ (where $x_q$ can be either $1/2$ or $1$ depending on the number of sites per unit cell),  so that
\begin{eqnarray}
\bar{Z}_n=Z_n \text{ mod } Z_q ,
\label{eq:zakbar}
\end{eqnarray}
where $Z_q=2\pi x_q=\pi$ or $2\pi$.

In the following, we consider several types of inversion-symmetric insulators and characterize them by their modified Zak phase.


\section{Inversion-symmetric insulators\label{sec:isi}}
\subsection{$\mathbb{Z}_2$ classification}
We define an angle $\vartheta$ characterizing one-dimensional (1D) band insulators by
\begin{eqnarray}
\vartheta = -2\pi \frac{\bar{Z}}{Z_q} \text{ mod } 2\pi,
\label{eq:angle}
\end{eqnarray}
where $\bar{Z}=\sum_{n <0} \bar{Z}_n$ is computed over the occupied bands~\cite{footnoteminus}. The difference between $\vartheta$ and $\bar{Z}$ is that $\vartheta$ is, by construction, defined mod $2\pi$. In the absence of a protecting symmetry, $\vartheta$ can take any value, so that there is a single class of 1D band insulators~\cite{Schnyder2008,Kitaev2009,Chiu2016}. 

However, for a crystal with inversion symmetry, the modified Zak phase must satisfy $\bar{Z}=-\bar{Z}$ mod $Z_q$ and therefore $\bar{Z}$ equals either $0$ or $Z_q/2$ mod $Z_q$~\cite{Zak1989, Zak1989b,Vanderbilt1993,Hughes2011,ChenLee2011}. This allows one to distinguish two classes of inversion-symmetric insulators depending on whether the angle
\begin{eqnarray}
\vartheta=0 \text{ (trivial) or } \pi  \text{ (topological)}.
\end{eqnarray}
The quantized angle $\vartheta$ plays the role of a $\mathbb{Z}_2$ topological invariant protected by inversion symmetry~\cite{footnotez2}. 

Any symmetry (not just inversion) that makes $\bar{Z}=-\bar{Z}$ mod $Z_q$ leads to these two classes~\cite{Qi2008}. It is conventional to call $\vartheta=0$ trivial and $\vartheta=\pi$ topological. The important thing is that there are two distinct classes, that can not be adiabatically connected and are distinguished by the value of the topological invariant $\vartheta$.

\subsection{Three types of inversion symmetry}
Depending on the position of the two inversion centers in the unit cell, we distinguish three types of inversion symmetries (see Fig.~\ref{fig:inversion}). The centers can be both mid-bond (bond inversion), both on-site (site inversion) or one on-site and the other mid-bond (mixed inversion). As representative examples of the three cases, we use two-band models:  the SSH$_{1/2}$ model (bond inversion), the CDW model (site inversion), and the Shockley model (mixed inversion)~\cite{Shockley}. In the following we study the three models in turn.

\begin{figure}
\begin{center}
\includegraphics[width=\linewidth]{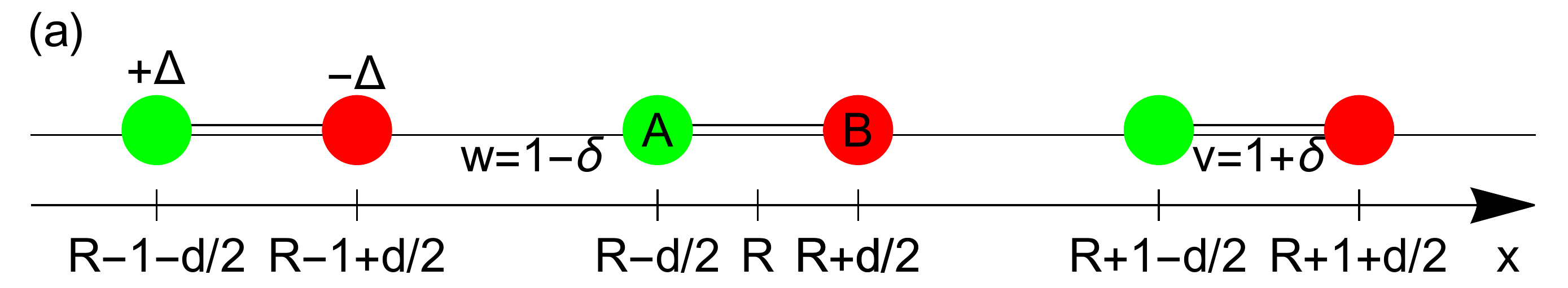} 
\includegraphics[width=\linewidth]{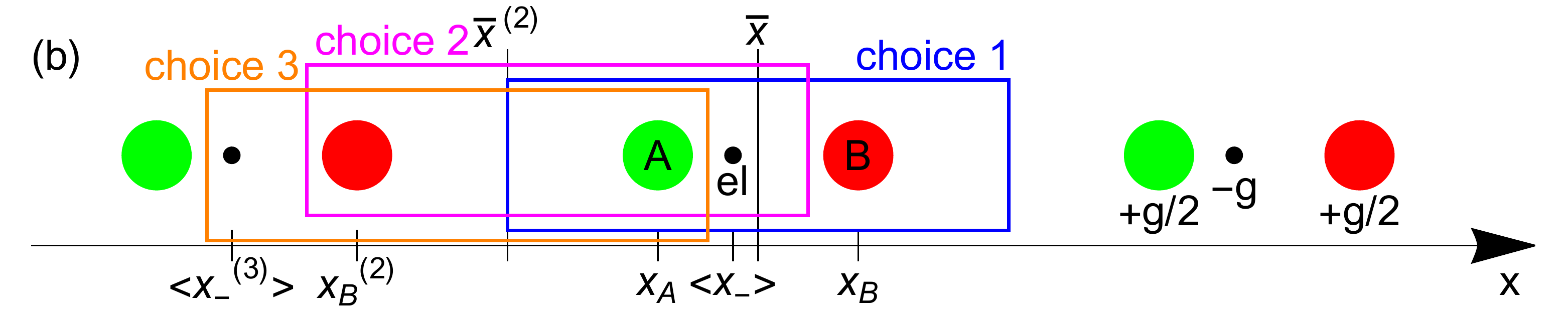} 
\caption{\label{fig:ucssh} (a) Rice-Mele model with arbitrary distance $d$ between $A$ and $B$ sites. The dimerization is called $\delta$ and the staggered on-site potential is called $\Delta$. SSH$_{1/2}$ corresponds to $\Delta=0$ and CDW to $\delta=0$. (b) Choices of unit cell. Electrons are localized at the Wannier center $\langle x_-\rangle$ shown as a black dot. The mean site position in a unit cell is $\bar{x}=(x_A+x_B)/2$. The electric charges $+g/2$ and $-g$, where $g=2$ is the spin degeneracy, refer to a specific ionic model discussed in App.~\ref{sec:ep}.
}
\end{center}
\end{figure}
%
%
%


\section{Su-Schrieffer-Heeger model\label{sec:ssh}}
\begin{figure}[h!]
\begin{center}
\includegraphics[width=0.49\linewidth]{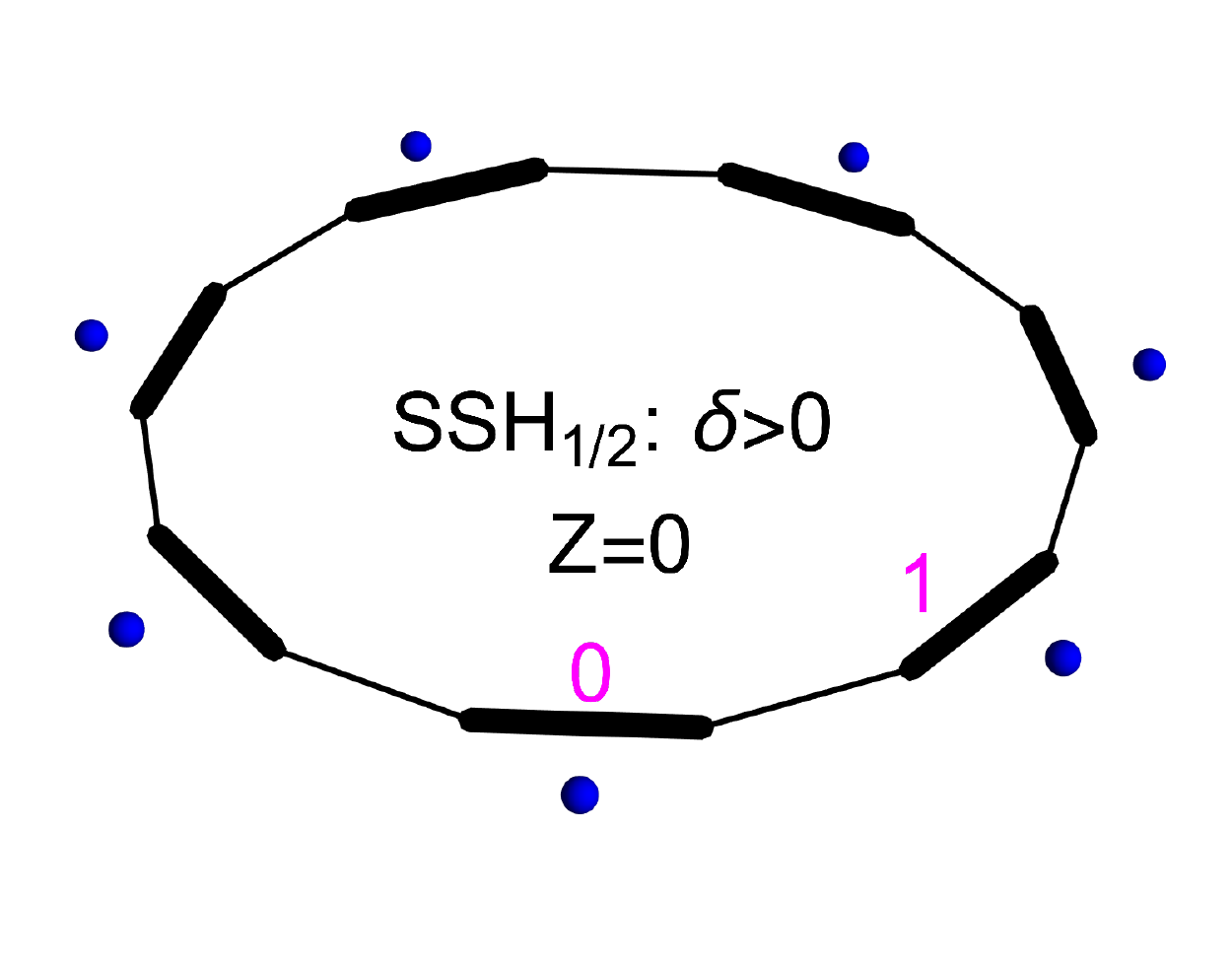} 
\includegraphics[width=0.49\linewidth]{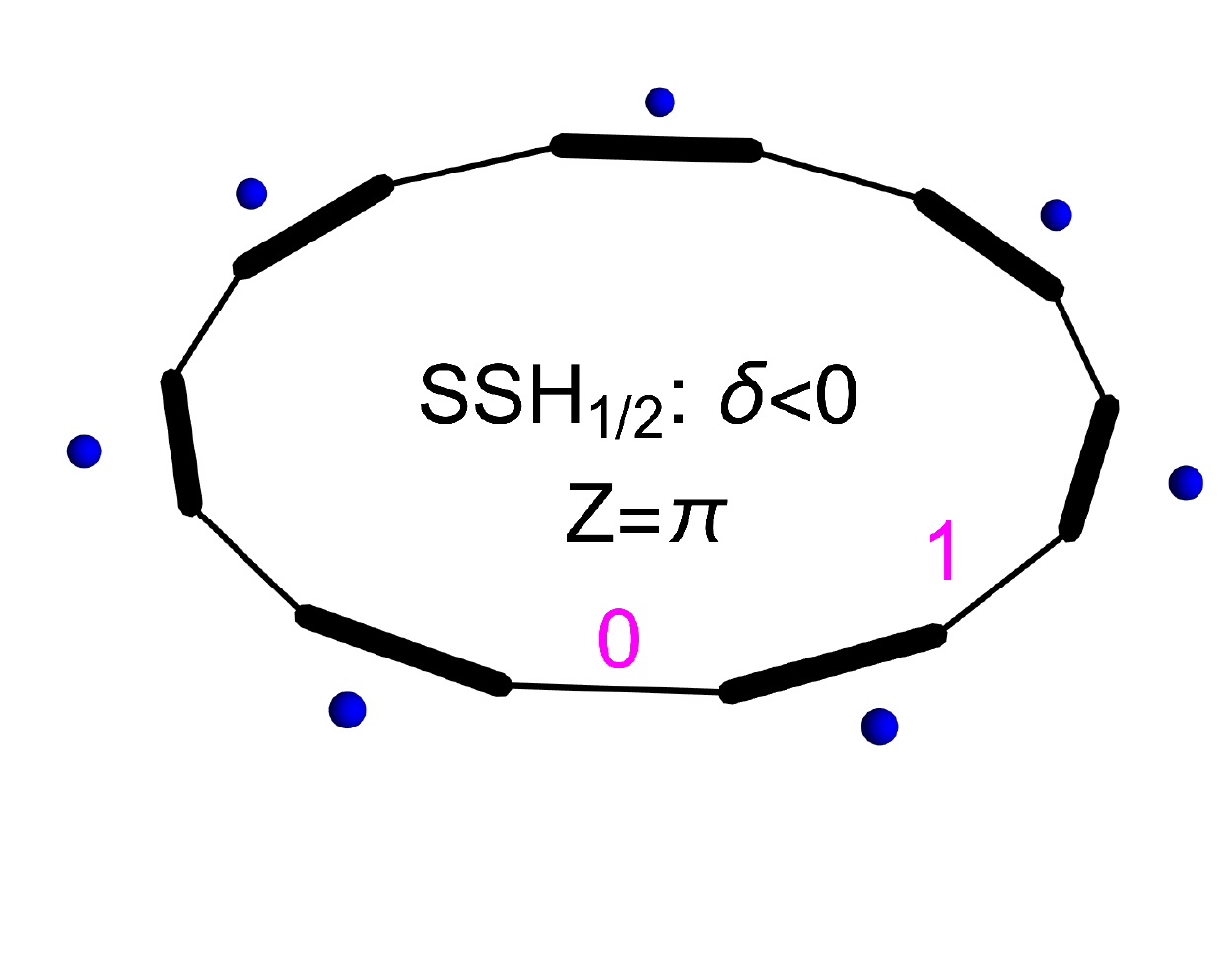} 
\caption{\label{fig:ssh1s2} The SSH$_{1/2}$ model on a ring: weak (strong) bonds are shown as thin (thick) lines, Wannier centers as blue dots and the position origin at an inversion center as a magenta 0. 
The lattice spacing is taken as unit length.
The two dimerizations are identical up to a change in the origin. 
The average position of sites in the unit cell is $\bar{x}=0$ mod $1/2$ so that $\bar{Z}$ matters mod $Z_q=\pi$. Therefore $\bar{Z}=0$ mod $\pi$.}
\end{center}
\end{figure}
The SSH model is a 1D tight-binding model with two sites $A$ and $B$ per unit cell, which was introduced to describe the valence electrons of trans-polyacetylene~\cite{SSH}. Each site $j=(R,l)$ contains a single $2p_z$ orbital $|j\rangle$, where $l=A,B$ labels the sublattices and $R$ is an integer that spans the Bravais lattice (with lattice spacing $a=1$). The two nearest-neighbor hopping amplitudes are $v=1+\delta$ (strong bond if the dimerization $\delta>0$) and $w=1-\delta$ (weak bond if $\delta>0$), following the pattern shown in Fig.~\ref{fig:ucssh}(a), with $\Delta=0$. The SSH model can be seen as a special case of the Rice-Mele model~\cite{RiceMele1982}, that we discuss in Appendix~\ref{sec:rm}.
The position operator reads
\begin{eqnarray}
x = \sum_{ j} x_{j} |j\rangle \langle j| ,
\label{eq:pos}
\end{eqnarray}
with $x_j=x_A=R-d/2$ if $j=(R,A)$ and $x_j=x_B=R+d/2$ if $j=(R,B)$, such that $\bar{x}=(x_A+x_B)/2=0$ mod $1/2$, see Fig.~\ref{fig:ucssh}(a). It is crucial that the two sites in the unit cell are not at the same position, i.e. $d\neq 0$, as they correspond to two different carbon atoms in polyacetylene. 
Physically, the strongest bond should have a shorter distance and $d$ should be a function of $\delta$, such as $d=1/2-\delta/4$. 
However, the model with $d=1/2$ for all $\delta$ already contains the essential features. Unless otherwise specified, we restrict ourselves to $d=1/2$ and call it SSH$_{1/2}$. The system has translation invariance: it is either infinite or finite with PBC.

The canonical Bloch Hamiltonian (see Appendix~\ref{sec:now}) reads
\begin{eqnarray}
H_d(k)
&=& 2\cos \frac{k}{2} \sigma_x^\text{eff}(k,d) - 2\delta \sin \frac{k}{2} \sigma_y^\text{eff}(k,d) ,
\label{eq:rmh}
\label{eq:hamII}
\end{eqnarray}
with the effective Pauli matrices 
\begin{eqnarray} 
\sigma_j^\text{eff} (k,d)&=&e^{i \phi \sigma_z/2} \sigma_j  e^{-i \phi \sigma_z/2}, 
\end{eqnarray}
where $\phi=k(d-1/2)$ and $j=x,y,z=1,2,3$. 
Note that $H_d(k+2\pi)\neq H_d(k)$. When $d=1/2$, we call $H(k)=H_{1/2}(k)$ and $\sigma_j^\text{eff}=\sigma_j $ are the standard Pauli matrices. 
The eigenvectors of (\ref{eq:rmh}) are cell-periodic Bloch states $|u_n(k)\rangle$ with band index $n=\pm$ and wavevector $k$ in the BZ $[-\pi, \pi[$. The corresponding eigenvalues do not depend on $d$ and read
\begin{eqnarray}
E_\pm(k) = \pm \sqrt{4\cos^2\frac{k}{2}+ 4 \delta^2 \sin^2 \frac{k}{2} }\, .
\label{eq:spec}
\end{eqnarray}

The SSH$_{d}$ model is time-reversal invariant $H_d(-k)^*=H_d(k)$, which implies that $E_n(-k)=E_n(k)$. It also has a pseudo-charge conjugation $\sigma_y H_d(-k)\sigma_y = - H_d(k)$, which implies that $E_-(k)=-E_+(-k)$. With time-reversal, it implies that $E_-(k)=-E_+(k)$. 

The two inversion centers per unit cell are mid-bond ($\bar{x}=(x_A+x_B)/2$ and $\bar{x}+1/2$). Bond-inversion symmetry acts as:
\begin{eqnarray}
H_d(k)\to \sigma_x H_d(-k)\sigma_x =H_d(k) \text{ when } \Delta=0\, .
\end{eqnarray}

The SSH$_d$ model is bipartite and there is a sublattice (chiral) symmetry, which means that:
\begin{eqnarray}
\sigma_z H_d(k)\sigma_z = - H_d(k)\, .
\end{eqnarray}
The corresponding sublattice pseudo-spin $\vec{\sigma}$ is not fully internal and is coupled to the intra-cell position because $d\neq 0$.

The SSH$_d$ model is invariant under the following parameter transformation $(d,\delta)\to (1-d,-\delta)$ and relabeling of the two sublattices ($A\leftrightarrow B$). 
On the Bloch Hamiltonian $H_{d,\delta}(k)$ in Eq.~(\ref{eq:hamII}), this symmetry acts as:
\begin{eqnarray}
H_{d,\delta}(k) \to \sigma_x H_{1-d,-\delta}(k) \sigma_x =  H_{d,\delta}(k)  \, .
\end{eqnarray}
Without loss of generality, one may therefore restrict the study to $\delta\geq0$. The SSH$_{1/2}$ model therefore has a single gapped phase, which is obvious from Fig.~\ref{fig:ssh1s2} (see also~\cite{ChenLee2011} and~\cite{Cooper2019}). In other words, it is possible to go from one dimerization ($\delta=\delta_0$) to the other ($\delta=-\delta_0$) without closing the gap, simply by a global shift of half a lattice spacing. The two dimerizations actually result from the spontaneous doubling of the lattice spacing at the Peierls transition. In analogy, the two possible symmetry-broken ground-states of an Ising anti-ferromagnet, which are also identical by half-a-lattice spacing shift, are not considered as two different phases. They just represent the two degenerate ground-states of the same anti-ferromagnetic phase. 

When $\delta>0$, the Wannier center of the occupied band is on the inversion center on the strong bond, i.e. $\langle x_- \rangle =0$ mod $1$ (see Fig.~\ref{fig:ucssh}) and $Z = 0$ mod $2\pi$. When $\delta<0$, the Wannier center is on the other inversion center, which is again on the strong bond, i.e. $\langle x_- \rangle =1/2$ mod $1$ (see Fig.~\ref{fig:ucssh}) and $Z = \pi$ mod $2\pi$. In the next paragraph, we show that $\bar{x}=0$ mod $1/2$ so that the modified Zak phase $\bar{Z}=0$ mod $\pi$.

In order to prove that $Z_q=\pi$, we consider several choices of unit cell and compute $\bar{Z}$, which should be independent of such a choice. In a given unit cell [see choice 1 in Fig.~\ref{fig:ucssh}(b)]
\begin{eqnarray}
\bar{Z}_1=2\pi (\langle x_- \rangle - \bar{x})\, ,
\end{eqnarray}
where $\bar{x}=(x_A+x_B)/2$ is the mean position of sites. In another choice [see 2 in Fig.~\ref{fig:ucssh}(b)], with the same $A$ ion, and the same Wannier center for the electrons, but another $B$ ion with position $x_B^{(2)}=x_B-1$, 
it reads $\bar{Z}_2 =\bar{Z}_1+\pi$ as $\bar{x}^{(2)}=\bar{x}-1/2$. If as a third choice  [see 3 in Fig.~\ref{fig:ucssh}(b)], we further move the unit cell to the left in order to change the Wannier center such as $\langle x_-^{(3)} \rangle=\langle x_- \rangle-1$, then $\bar{Z}_3 =\bar{Z}_2-2\pi=\bar{Z}_1-\pi$. Other choices always lead to modified Zak phases that differ by an integer multiple of $\pi$, so that $Z_q=\pi$ and not $2\pi$. 

This shows that the two dimerizations of SSH$_{1/2}$ describe the same phase with $\bar{Z}=0$ mod $\pi$ (see Fig.~\ref{fig:ssh1s2}). The conclusion is that there is no phase transition as $\delta$ changes sign, despite the gap closing, and that the gapped phase is characterized by $\vartheta=0$.


\section{Charge density wave model\label{sec:cdw}}
\begin{figure}[h!]
\begin{center}
\includegraphics[width=0.49\linewidth]{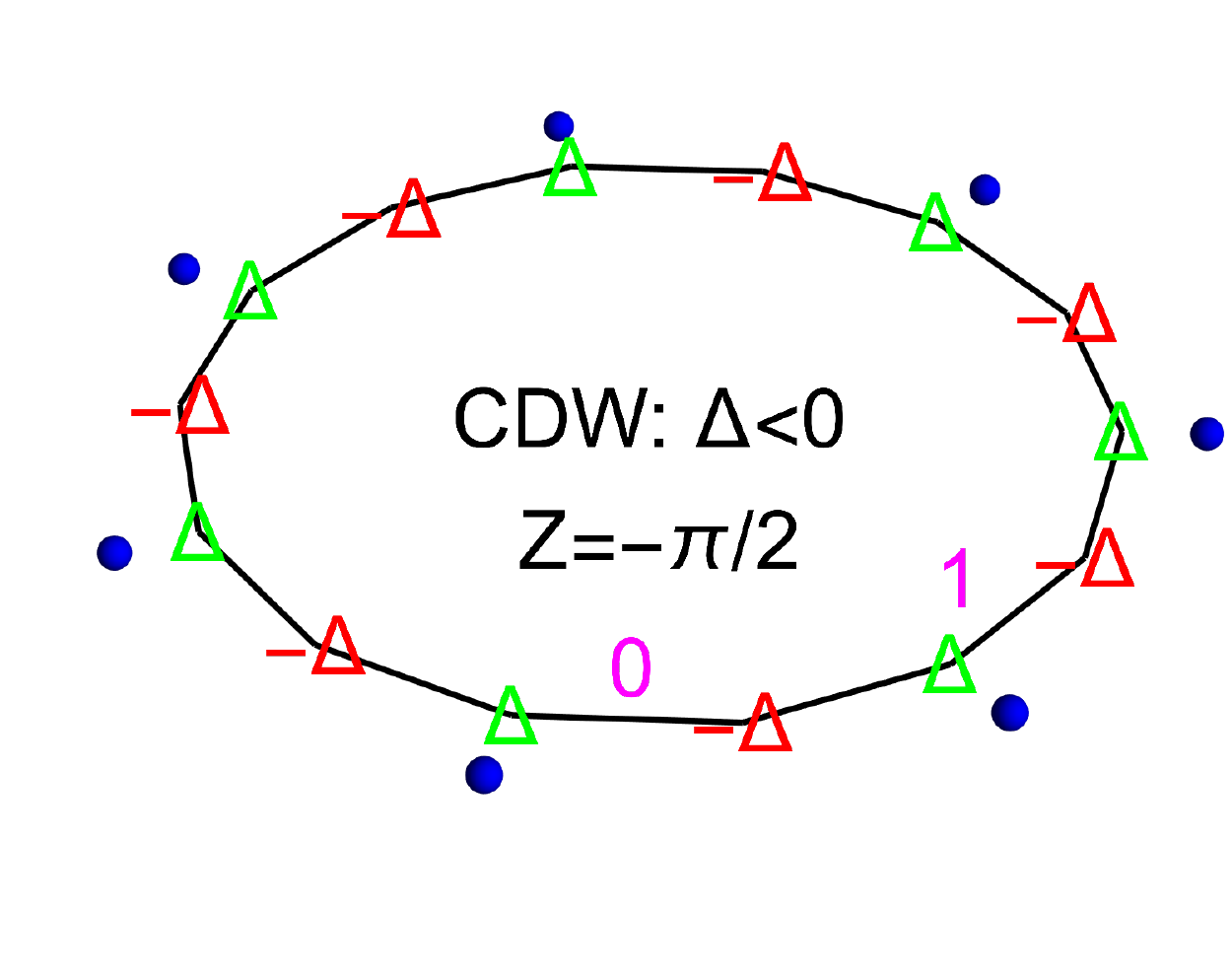} 
\includegraphics[width=0.49\linewidth]{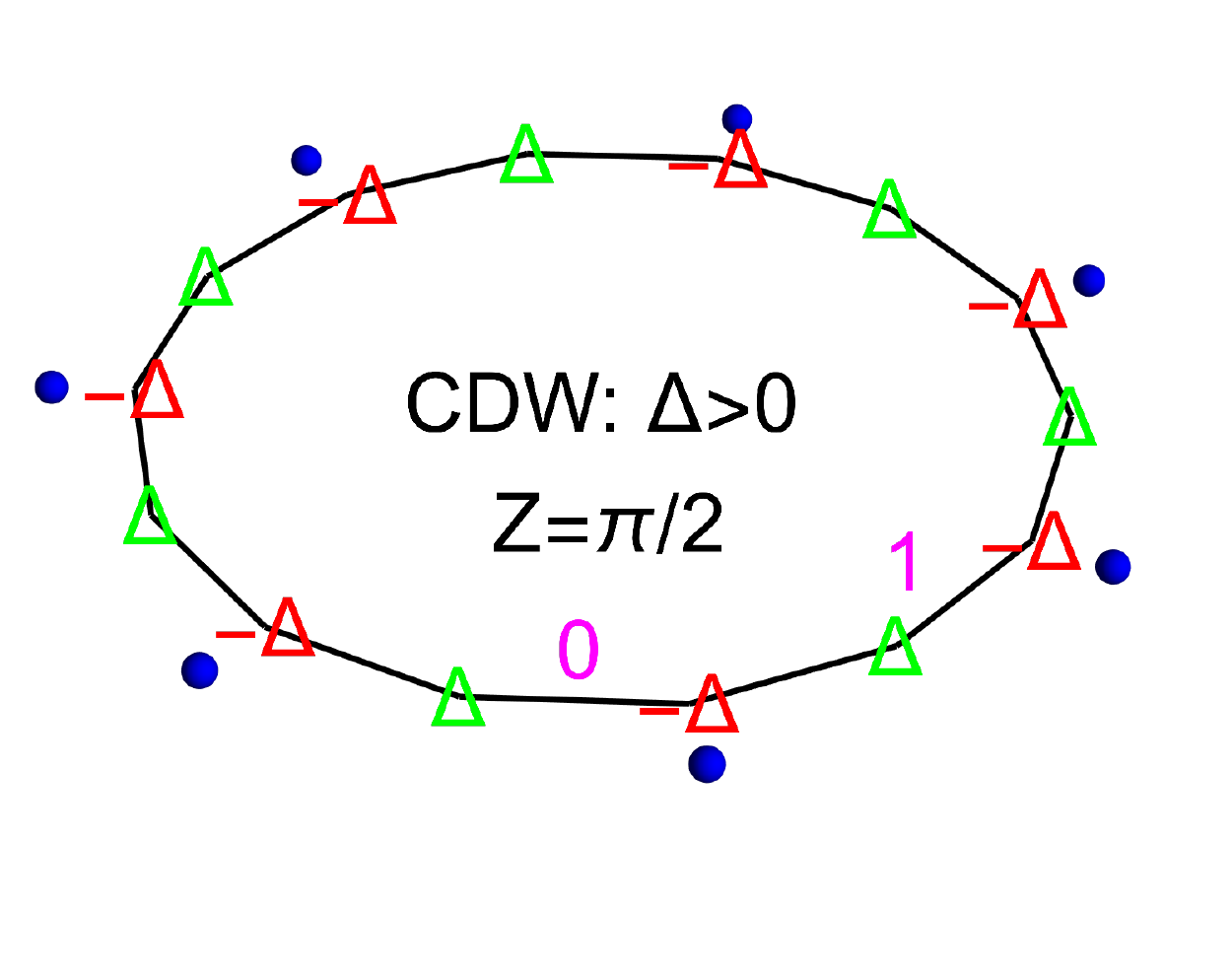} 
\caption{\label{fig:cdw}The CDW model on a ring: the staggered on-site potential $\pm \Delta$ is shown in green/red, Wannier centers as blue dots and the position origin as a magenta 0. 
The two staggerings are identical up to a change in the origin. 
The average position of sites in the unit cell is $\bar{x}=0$ mod $1/2$ so that $\bar{Z}$ matters mod $Z_q=\pi$. Therefore $\bar{Z}=\pi/2$ mod $\pi$.}
\end{center}
\end{figure}
Another model of a 1D two-band inversion-symmetric insulator is the CDW. It also has two sites per unit cell, but instead of dimerized hoppings, it has a staggered on-site potential $\Delta$ (see Fig.~\ref{fig:ucssh}(a) with $\delta=0$). It can be seen as a special case of the Rice-Mele model~\cite{RiceMele1982}, see Appendix~\ref{sec:rm}. Inversion symmetry is of site type and only exist for an $A$-$B$ distance $d=1/2$. The canonical Bloch Hamiltonian (see Appendix~\ref{sec:now}) reads
\begin{eqnarray}
H(k)
&=& 2\cos \frac{k}{2} \sigma_x + \Delta \sigma_z ,
\label{eq:hamcdw}
\end{eqnarray}
which is also not BZ-periodic as $H(k+2\pi)\neq H(k)$.

Inversion acts as
\begin{eqnarray}
H(k)\to H(-k) =H(k) \, ,
\end{eqnarray}
and  the two inversion centers are on-site ($x_A$ and $x_B$).  

The CDW model is not bipartite but there is another chiral symmetry:
\begin{eqnarray}
\sigma_y H(k)\sigma_y = - H(k)\, .
\end{eqnarray}

The model is also invariant under the following parameter transformation $\Delta\to -\Delta$ and relabeling of the two sublattices ($A\leftrightarrow B$). 
On the Bloch Hamiltonian $H_{\Delta}(k)$ in Eq.~(\ref{eq:hamcdw}), this symmetry acts as:
\begin{eqnarray}
H_{\Delta}(k) \to \sigma_x H_{-\Delta}(k) \sigma_x =  H_{\Delta}(k)  \, .
\end{eqnarray}
Without loss of generality, one may therefore restrict the study to $\Delta\geq0$. 

When $\Delta>0$, the Wannier centers of the occupied band are localized on the $B$ ions that have lowest on-site energy and $\langle x_-\rangle = 1/4$ mod $1$ so that $Z=-\pi/2$, see Fig.~\ref{fig:cdw}. When $\Delta <0$, they are on the $A$ ions that have lowest energy $\langle x_-\rangle = -1/4$ mod $1$, so that $Z=\pi/2$. Because the mean site position $\bar{x}=0$ mod $1/2$, due to the two sites per unit cell, we find that $Z_q=\pi$ and that the modified Zak phase $\bar{Z}=\frac{\pi}{2}$ mod $\pi$.

Here, as for SSH$_{1/2}$, there is a single gapped phase and no phase transition as $\Delta$ changes sign. However, the gapped phase is such that $\vartheta=\pi$ and represents the other class of inversion-symmetric insulators. 

It is easy to construct an inversion-symmetric model that features a phase transition between SSH$_{1/2}$ and CDW, see~\cite{Cayssol2021}. Below we study a more physical example that features a topological phase transition between $\vartheta=0$ and $\pi$.


\section{Shockley model\label{sec:shockley}}
\begin{figure}[h!]
\begin{center}
\includegraphics[width=0.49\linewidth]{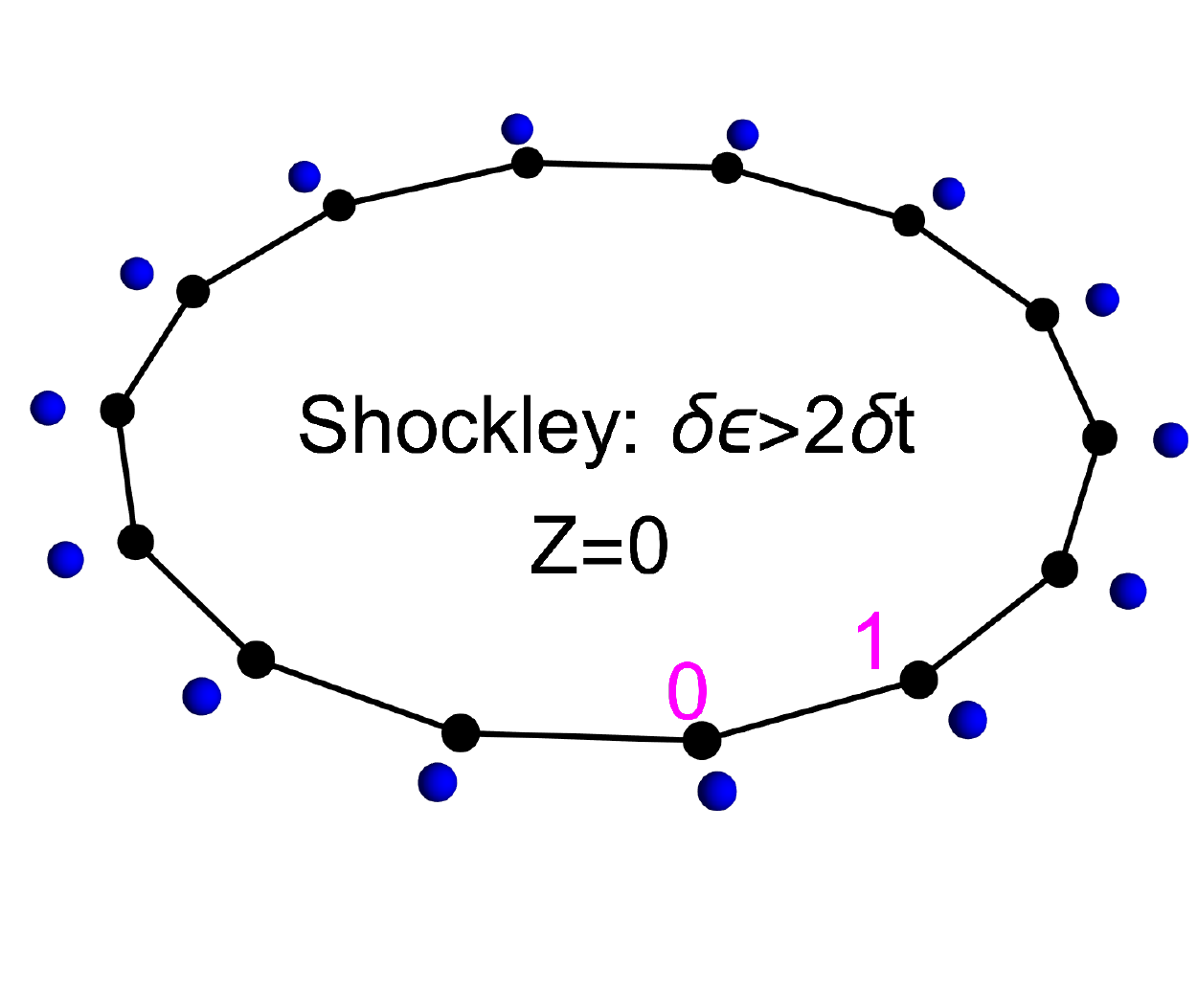} 
\includegraphics[width=0.49\linewidth]{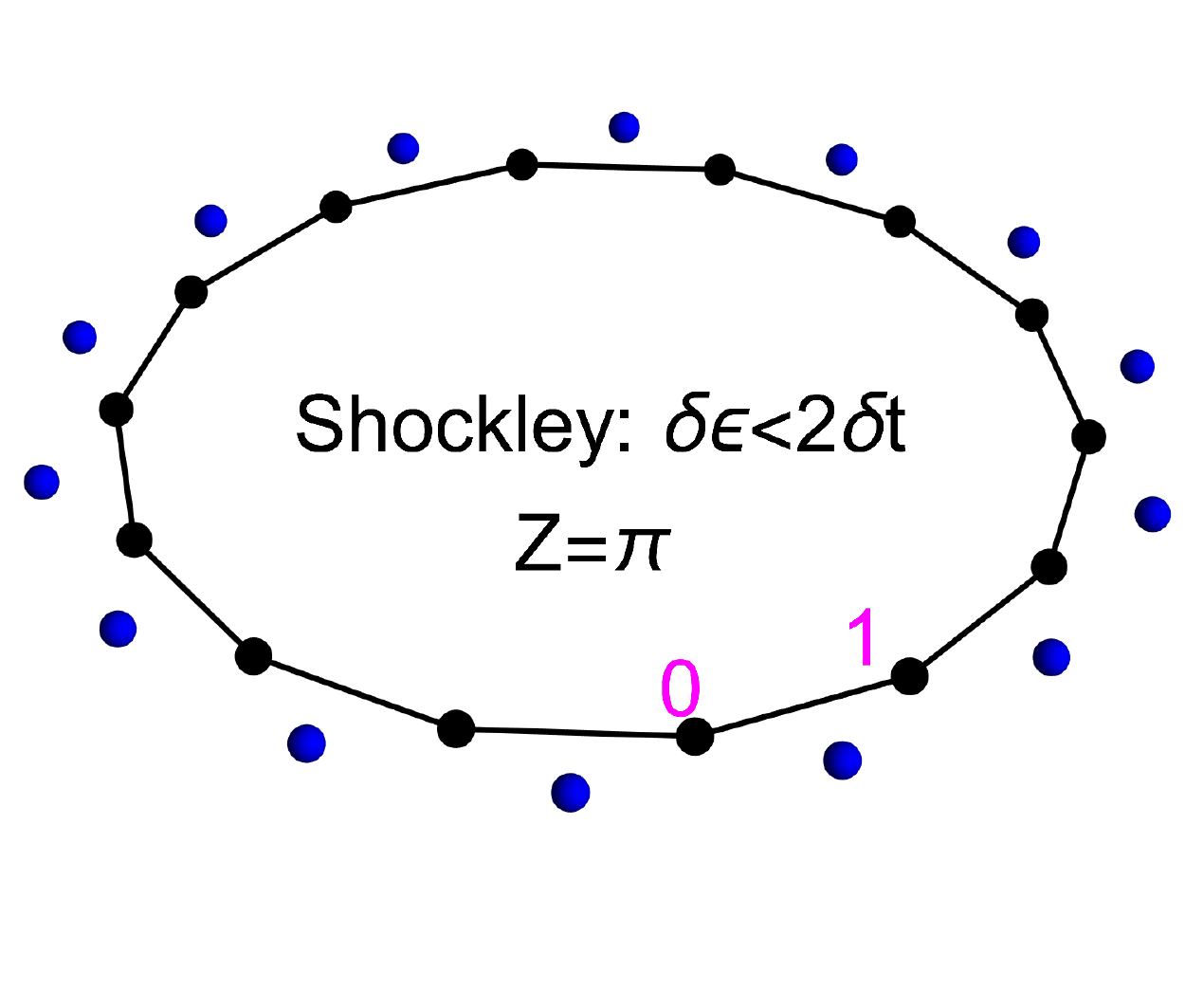} 
\caption{\label{fig:shockley} 
The Shockley model on a ring: sites are shown as black dots, Wannier centers as blue dots and the position origin as a magenta 0. The average position of sites in the unit cell is $\bar{x}=0$ mod $1$ so that $\bar{Z}$ matters mod $Z_q=2\pi$. Therefore $\bar{Z}=\pi \Theta(2 \delta t - \delta \epsilon)$ mod $2\pi$.}
\end{center}
\end{figure}
The Shockley model of coupled $s$ and $p$ bands~\cite{Vanderbilt1993,Shockley} is sometimes assumed to be equivalent to SSH (see e.g. ~\cite{Yakovenko2012} and the Supp. Mat. in \cite{,Bradlyn2017}). Here, we show that it is different from SSH$_{1/2}$ but closely related to SSH$_0$. In particular, due to a single site per unit cell, $Z_q=2\pi$ and its canonical Bloch Hamiltonian is $2\pi$-periodic (see Fig.~\ref{fig:shockley}).

We follow the analysis of~\cite{Vanderbilt1993}. The Shockley model is a 1D tight-binding model with a single site per unit cell and two orbitals ($s$ and $p_x$). The $s$ ($p_x$) orbital is even (odd) under inversion. The crucial difference with SSH$_{1/2}$ and CDW is that there is a single site per unit cell rather than two. This means that here $Z_q=2\pi$ as $\bar{x}=0$ mod $1$.

The Bloch Hamiltonian is
\begin{eqnarray}
H(k)&=&[\frac{\epsilon_s+\epsilon_p}{2}+(t_{ss}+t_{pp})\cos k ]\sigma_0   \\
&+& [ \frac{\epsilon_s-\epsilon_p}{2}+(t_{ss}-t_{pp})\cos k ] \sigma_z + 2 t_{sp} \sin k \, \sigma_y\, , \nn
\end{eqnarray}
where $t_{ss}\leq 0$, $t_{pp}\geq 0$ and $t_{sp}$ are hopping amplitudes and $\epsilon_s\leq 0$ and $\epsilon_p\geq 0$ are orbital energies. Mixed inversion symmetry means $\sigma_z H(-k) \sigma_z =H(k)$.

We call $\delta \epsilon = \epsilon_p - \epsilon_s$ and $\delta t = t_{pp}-t_{ss}$. When $\delta \epsilon > 2 \delta t$, the Wannier center is on-site $\langle x_-\rangle =0$ mod $1$, so that $\bar{Z}=0$ mod $2\pi$ and the system is an atomic insulator with $s$-type valence band. A band inversion occurs at $\delta \epsilon = 2 \delta t$, where the gap closes. When $\delta \epsilon < 2 \delta t$, the Wannier center is mid-bond $\langle x_- \rangle =1/2$ mod $1$ and the modified Zak phase $\bar{Z}=\pi$ mod $2\pi$. It is a covalent insulator with a valence band of bonding  $sp$-hybridized type. Here there is a genuine topological phase transition between a phase with $\vartheta=0$ and one with $\vartheta=\pi$ as $\delta \epsilon - 2 \delta t$ changes sign.

Generically, the Shockley model does not have a chiral symmetry, belongs to the AI class and is therefore a trivial insulator according to the periodic table~\cite{Schnyder2008,Kitaev2009,Chiu2016}. But it still has a $\mathbb{Z}_2$ classification as an inversion-symmetric insulator. The fact that there is no chiral symmetry (and hence no winding number) does not change anything for the quantization of the modified Zak phase.

However, for $\epsilon_s=-\epsilon_p$ and $t_{ss}=-t_{pp}$, the model possesses a chiral (not sublattice) symmetry 
\begin{eqnarray}
\sigma_x H(k)\sigma_x = - H(k)\, ,
\end{eqnarray}
and belongs to the BDI class~\cite{Schnyder2008,Kitaev2009,Chiu2016}. In contrast to the SSH$_{1/2}$ model, here the pseudo-spin $\vec{\sigma}$ is internal because the two orbitals are at the same position. As the canonical Bloch Hamiltonian $H(k)$ is $2\pi$-periodic, unlike for SSH$_{1/2}$ and CDW, there is a well-defined winding number $W=\text{sign } t_{sp}$ if $\delta \epsilon < 2 \delta t$ and $W=0$ if  $\delta \epsilon > 2 \delta t$. The winding is related to the modified Zak phase by:
\begin{eqnarray}
\bar{Z}=\pi W \text{ mod } 2\pi \, .
\end{eqnarray}
Only in this very specific case, does the winding number and the modified Zak phase coincide (up to a factor of $\pi$). But this is the exception rather than the rule~\cite{footnotezakwinding}. This is how the $\mathbb{Z}$ classification for chiral insulators (with invariant $W$) becomes a $\mathbb{Z}_2$ classification for inversion-symmetric insulators (with invariant $\bar{Z}$ or $\vartheta$) and $W$ matters only mod 2. 
\begin{figure}[h!]
\begin{center}
\includegraphics[width=0.49\linewidth]{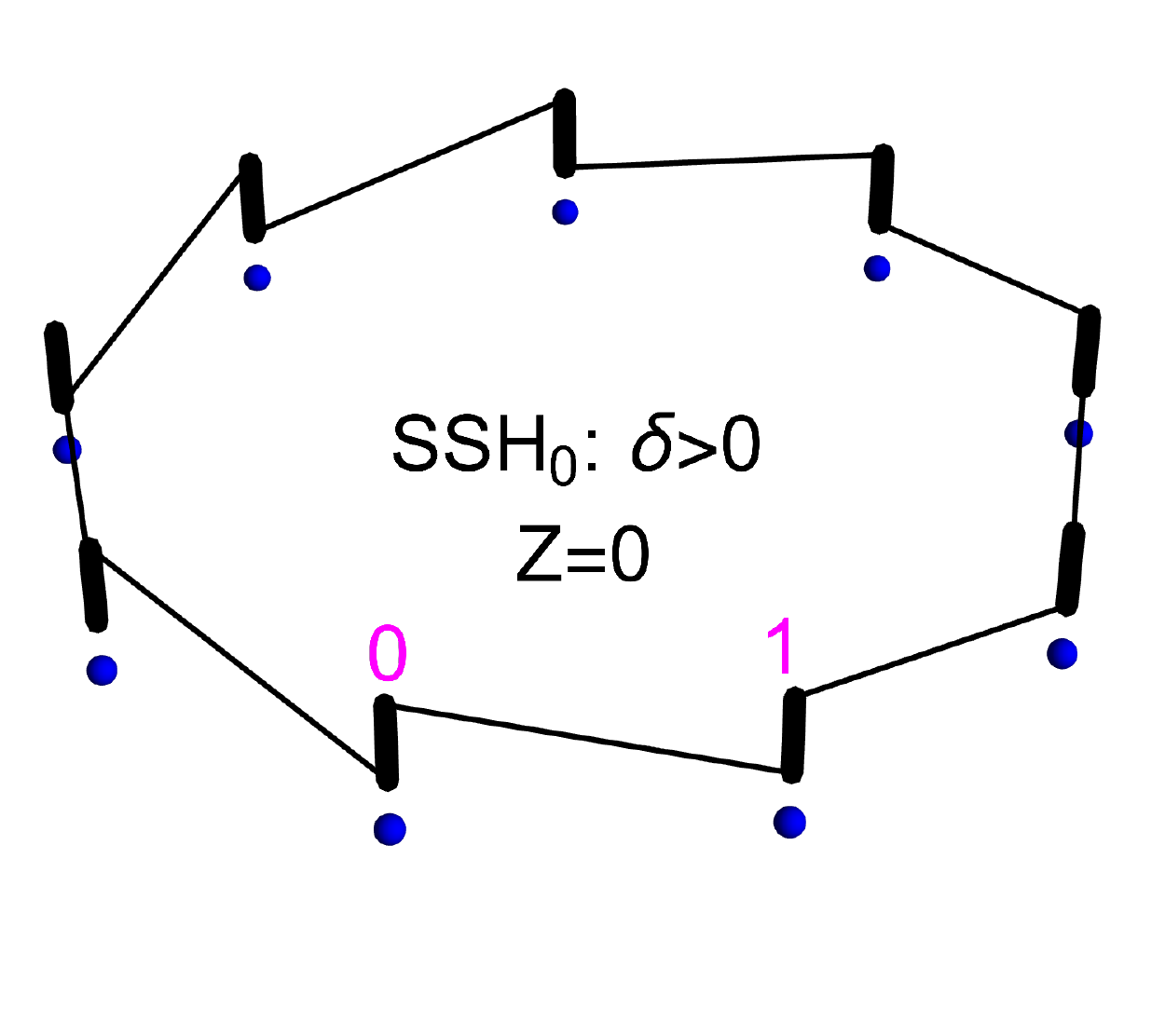} 
\includegraphics[width=0.49\linewidth]{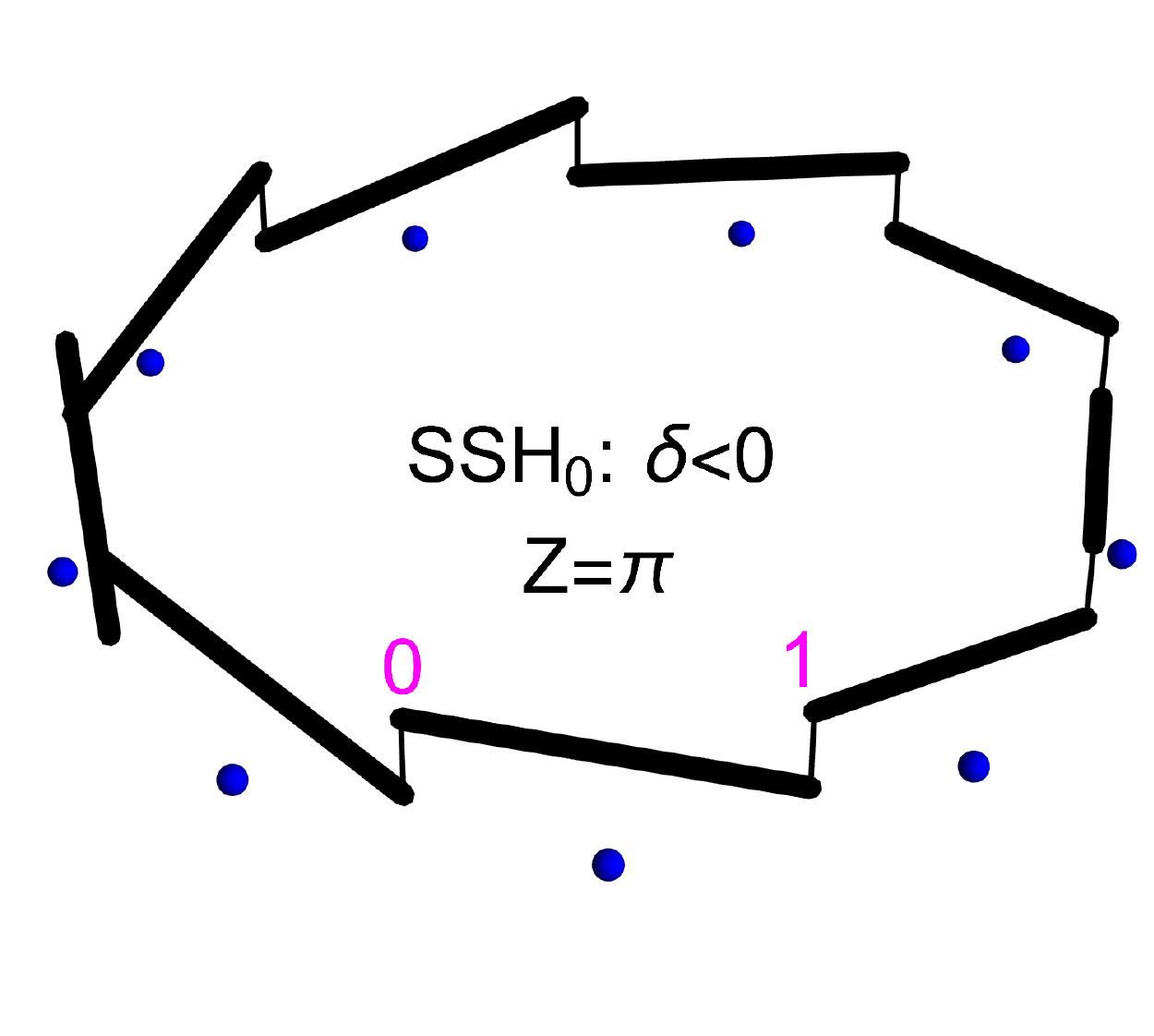} 
\caption{\label{fig:ssh0} The SSH$_{0}$ model on a ring. The $B$ orbital is on top of the $A$ one (vertical direction is just for visualization purposes). Here $Z_q=2\pi$ so that $\bar{Z}=\pi \Theta(-\delta)$ mod $2\pi$.}
\end{center}
\end{figure}

The SSH$_0$ model has mixed inversion and can be mapped onto the chiral Shockley model (see Fig.~\ref{fig:ssh0}). Other models that have a similar behavior are the Creutz ladder at $\pi/2$-flux per plaquette~\cite{Creutz1999} and the Kitaev chain of a $p$-like superconductor~\cite{Kitaev2001}. In these four models (Shockley, SSH$_0$, Creutz, Kitaev), the two ``orbitals" are at the same $x$-position. This is the key difference with the SSH$_{1/2}$ model.


\section{Conclusion\label{sec:ccl}}
We studied three types of inversion-symmetric insulators and our main results are summarized in Table~\ref{table}. The gap closings that occur in these three models are of different nature. In all cases, at the gap closing, the inversion symmetry is of mixed type. In other words, only for the Shockley model does the inversion symmetry remain of the same type across the transition. 

We find that SSH$_{1/2}$ (resp. CDW) has a unique gapped phase with $\vartheta=0$ (resp. $\pi$), which makes it a trivial (resp. topological) inversion-symmetric insulator. There is no topological phase transition as the dimerization $\delta$ (resp. the on-site potential $\Delta$) changes sign.

In contrast, the Shockley model is arguably the simplest example of an inversion-symmetric insulator featuring a topological transition between a phase with $\vartheta=0$ and one with $\vartheta=\pi$. In a special limit, it also provides an example of topological phase transition in a chiral insulator characterized by a winding number.

As chiral insulators, SSH$_{1/2}$ and CDW do not have a well-defined bulk winding number, see Appendix~\ref{sec:now} and also~\cite{Guzman2020}. This is due to the two sites per unit cell, which prevent the existence of a BZ-periodic Bloch Hamiltonian that is independent of a unit cell choice. 

Breaking chiral (but not inversion) symmetry does not change the topological invariant $\vartheta$. Sublattice (chiral) is not an exact symmetry of real polyacetylene, in contrast to inversion symmetry, as it is broken by next-nearest neighbor hopping. The SSH$_{1/2}$ model, which has both symmetries, is better characterized as a trivial inversion-symmetric insulator than as a chiral insulator.

\begin{widetext}
\begin{center}
\begin{table}
\begin{tabular}{c||c|c||c|c||c|c}
model & SSH$_{1/2}$ & SSH$_{1/2}$ & CDW & CDW & Shockley & Shockley \\
\hline
\hline
phase & $\delta> 0$ & $\delta< 0$ & $\Delta> 0$ & $\Delta< 0$  & $\delta \epsilon > 2 \delta t $  & $\delta \epsilon < 2 \delta t $ \\
\hline
inversion centers & mid-bonds & mid-bonds & on-sites & on-sites & mixed & mixed \\
\hline
position of sites & 2c & 2c  &1a and 1b &1a and 1b & 1a & 1a \\
\hline
mean site position $\bar{x}$ & $0$ mod $1/2$ & $0$ mod $1/2$  & $1/4$ mod $1/2$ & $1/4$ mod $1/2$ & $0$ mod $1$ & $0$ mod $1$ \\
\hline
Wannier centers & mid-bond  & mid-bond & on-site & on-site  & on-site  & mid-bond \\
(position of electrons) & 1a & 1b & 1b & 1a & 1a & 1b \\
\hline
$Z_q$ & $\pi$ & $\pi$ & $\pi$ & $\pi$  & $2\pi$  & $2\pi$ \\
\hline
$\bar{Z}$ mod $Z_q$ & $0$ & $0$ & $Z_q/2=\pi/2$ &  $Z_q/2=\pi/2$ & $0$  &$Z_q/2=\pi$ \\
\hline
$\vartheta$ mod $2\pi$& $0$  & $0$  & $\pi$ & $\pi$ & $0$ &$\pi$ \\
\hline
bulk winding $W$ & undefined & undefined & undefined & undefined & $^{(*)}$ $0$ & $^{(*)}$ $\text{sign }t_{sp}$ \\
\hline
insulator type & molecular  & molecular & ionic & ionic & atomic & covalent \\
\end{tabular}
\caption{Three types of inversion-symmetric insulators: SSH$_{1/2}$, CDW and Shockley. 1a, 1b and 2c refer to Wyckoff positions in the unit cell (see e.g. Fig.~S5 in the Supp. Mat. of~\cite{Bradlyn2017}).  The two inversion centers are always at 1a and 1b, but can be on-site, mid-bond, or mixed. The position origin is chosen on an 1a inversion center. At the transition, the gap vanishes and inversion symmetry is of mixed type in all three cases. $^{(*)}$Defined only for the chiral Shockley model (see Sec.~\ref{sec:shockley}).
\label{table}}
\end{table}
\end{center}
\end{widetext}

Although the SSH$_{1/2}$ chain is not a topological insulator, it is a Dirac insulator, with the property of trapping mid-gap states at domain walls~\cite{Jackiw1976,SSH}. Our conclusion does not contradict the bulk-edge correspondence of chiral insulators in terms of an edge-dependent (relative) winding number~\cite{Ryu2002,Delplace2011,footnotezakwinding,Guzman2020}. 

The quantum of Zak phase $Z_q$ depends on the number of sites per unit cell. The Shockley model has a single site per unit cell and $Z_q=2\pi$. However, the CDW and SSH$_{1/2}$ models have two sites per unit cell and $Z_q=\pi$. 

Because they do not involve the electric charge (e.g. a model for ions) but only the position of the orbitals in the tight-binding description, our results for the modified Zak phase also apply to artificial realizations of SSH$_{1/2}$ with ``neutral electrons'' such as cold atoms in optical lattices~\cite{Atala2013}, microwaves in lattices of dielectric resonators~\cite{Dutreix2020} or polaritons~\cite{StJean2017}. 

In contrast to 2D topological invariants (e.g. the Chern number~\cite{footnotecurvature}), the 1D topological invariant $\vartheta$ (or $\bar{Z}$) depends not only on the tight-binding Hamiltonian $H$ but also on the position operator $x$. This is due to the fact that the Zak phase explicitly depends on the position operator. In other words, the mere knowledge of the connectivity between orbitals, contained in $H$, is not enough to decide of the topology of 1D bands and one must know their orbital embedding, contained in $x$. The topological classification of 1D band insulators~\cite{Schnyder2008,Kitaev2009,Chiu2016} assumes a large number of bands and stability to the addition of trivial bands (the so-called stable equivalence). It does not necessarily agree with a specific two-band model. For example, it agrees with SSH$_0$ but not with SSH$_{1/2}$ (actually, it does not distinguish the two). The SSH$_0$ model, which is very unphysical in describing polyacetylene as its strongest bond is not always the shortest, has a well-defined bulk winding number and is equivalent to the chiral Shockley model.

As perspectives, we would like to study whether the orbital embedding is also crucial for inversion-symmetric insulators with more than two bands. For example, by adding orbitals, would it be possible to continuously connect the SSH$_{1/2}$ and SSH$_0$ models? Also, in the case of an insulator with more than two bands, what are the possible values of the quantum of Zak phase $Z_q$ depending on the number of filled bands?

\section*{Acknowledgements}
We thank J. Asb\'oth, M. Ayachi, J. Cayssol, J. Kellendonk, A. Mesaros, D. Roy, C. Tauber, L. Trifunovic and J. Vidal for useful discussions.

\begin{appendix}

\section{Canonical versus periodic Bloch Hamiltonian: influence on the winding number \label{sec:now}}
The SSH$_{1/2}$ model has a chiral sublattice symmetry, which means that its $2\times 2$ Bloch Hamiltonian can be represented as an effective magnetic field pointing along the equator of the Bloch sphere. This defines a map from the BZ circle to the equator, which should be characterized by a winding number. Then why do we say that there is no well-defined bulk winding number? 

This has to do with the different conventions for the Bloch Hamiltonian when there are several sites per unit cell, see e.g.~\cite{Bena2009,Fuchs2010,Fruchart2014,Lim2015,Vanderbilt2018}. When going from the tight-binding Hamiltonian $H$ to the $k$-dependent Bloch Hamiltonian, one needs to define the Fourier transform. Because of the two sites per unit cell, there are several ways of doing this, sometimes called basis I and basis II~\cite{Bena2009} (or convention II and convention I~\cite{Vanderbilt2018})  in the literature. 

One way leads to a periodic Bloch Hamiltonian $\mathcal{H}(k)$, i.e. such that $\mathcal{H}(k+2\pi)=\mathcal{H}(k)$, which involves the unitary transformation $e^{-i k R} H e^{i k R}$, where $R$ is the Bravais lattice position operator (basis I). However, this way is not unique as it depends on the choice of unit cell. Each choice of unit cell gives another periodic Bloch Hamiltonian and another value for the winding number, which is therefore not physical. 

The other way leads to a unique Bloch Hamiltonian $H(k)$ (which we call canonical, or basis II) independent of a unit cell choice, which involves the unitary transformation $e^{-i k x} H e^{i k x}$, where $x$ is the complete position operator. However it is not periodic, i.e. $H(k+2\pi)=e^{-i2\pi(x-R)}H(k)e^{i 2\pi (x-R)}\neq H(k)$, as it depends on the distance $d$ between the two sublattices, so that a winding number can not be defined.

The conclusion is that there is no well-defined bulk winding number for the SSH$_{1/2}$ model. The same issue applies to the CDW model. However, the chiral Shockley model and the SSH$_0$ model do not suffer from this problem as they have a single site per unit cell.


\section{Electric polarization\label{sec:ep}}
 According to the modern theory of polarization, the Zak phase gives the electronic contribution to the bulk electric polarization of crystals~\cite{King-Smith1993,Vanderbilt1993,Resta2007}. As it is only well-defined for a neutral system, the polarization also involves the ions. One must therefore specify an ionic model, e.g. by giving the position and charge of static point-like ions. 
 
There is a specific ionic model that is such that the polarization is proportional to the modified Zak phase $\bar{Z}$. It consists in assuming that every site of the electronic (tight-binding) model carries an ion and that all ions have the same charge. This ionic model is as harmless as possible to the electronic model. It is actually the only model compatible with polyacetylene. For SSH$_{1/2}$ and CDW (and more generally Rice-Mele, see~\cite{Vanderbilt1993}), if we consider electrons of charge $-1$ each and spin degeneracy $g=2$, this leads to a $g/2=1$ ion charge on every site, see Fig.~\ref{fig:ucssh}(b). For the Shockley model, this means a charge $2$ ion on each site. With this specific ionic model, in the three cases (SSH$_{1/2}$, CDW and Shockley), the polarization is: 
\begin{eqnarray}
P= -g \frac{Z}{2\pi} + g \bar{x}=-g \frac{\bar{Z}}{2\pi}\, \text{ mod } P_q,
\label{eq:bep}
\end{eqnarray}
where $P_q$ is the quantum of polarization. Usually, $P_q=g$ which reflects the mod $2\pi$ of the Zak phase $Z$~\cite{King-Smith1993,Vanderbilt1993,Resta2007}. This is the case of the Shockley model, because $\bar{x}=0$ mod $1$. However, when $g=2$ and there are ions of odd-integer charge (as for SSH$_{1/2}$ or CDW), it is known that $P_q=1$ and not $2$, see page 169 in~\cite{Vanderbilt2018} and~\cite{Kudin2007}. It turns out that for SSH$_{1/2}$ and CDW, $P_q=g/2$ because $\bar{x}=0$ mod $1/2$, i.e. for the same reason that $Z_q=\pi$. Therefore $P=0$ mod $g/2$ for SSH$_{1/2}$ and $P=g/4$ mod $g/2$ for CDW. 

For this specific ionic model, the quantity
\begin{eqnarray}
\vartheta = 2\pi \frac{P}{P_q} \text{ mod } 2\pi
\end{eqnarray}
is the well-known polarization angle appearing in the electromagnetic Lagrangian of 1D dielectrics~\cite{ChenLee2011}. For other ionic models, $P$ is no longer directly proportional to $\bar{Z}$.

To conclude this section on electric polarization, we show that the bulk polarization and the polarization quantum are not a mere convention but are measurable in principle via the end charge $Q$ of an open chain. The surface charge theorem~\cite{Vanderbilt1993} relates $Q$ to the bulk polarization $P$ by
\begin{eqnarray}
Q=P+mP_q,
\end{eqnarray}
where $m$ is an integer. The end charge of a finite chain is obviously a measurable quantity. Changing the terminations of an open chain, it is possible to change the end charge by integers $\Delta Q= \Delta m P_q \in \mathbb{Z}$ corresponding to the transfer of individual electrons from one end to the other. For a chain with bulk inversion symmetry, $P=0$ or $P_q/2$ mod $P_q$ and therefore the end charge $Q=m P_q$ or $(m+1/2)P_q$. By monitoring the allowed values of $Q$ for a chain with inversion symmetry as a function of different surface termination, one may determine $P$ and $P_q$. 

For example, Vanderbilt and King-Smith (see Fig.~4 in~\cite{Vanderbilt1993}) have computed the end charge as a function of a parameter (called $\theta$) in the spinful Rice-Mele model (see Appendix~\ref{sec:rm}). For the two values of this parameter ($\theta=\pi/2$ and $3\pi/2$) that correspond to the two dimerizations of SSH$_{1/2}$, they find $Q=1$ and $0$, which means that $P=0$ and that $P_q=1/m$ with $m$ a positive integer. For the two values of the parameter ($\theta=0$ and $\pi$)  that correspond to the two versions of CDW, they find $Q=-1/2$ and $1/2$, which means that $P_q=1$ and $P=1/2$. The conclusion is again that $P_q=g/2$ when $g=2$, that $P=0$ mod $P_q$ for SSH$_{1/2}$ and that $P=P_q/2$ mod $P_q$ for CDW.


\section{Rice-Mele model \label{sec:rm}}
In this appendix, we generalize the SSH$_{1/2}$ and CDW models to the Rice-Mele (RM) model~\cite{RiceMele1982}. The latter has both a dimerization $\delta$ and a staggered on-site potential $\Delta$, see Fig.~\ref{fig:ucssh}(a). The canonical Bloch Hamiltonian  reads
\begin{eqnarray}
H_{d,\Delta,\delta}(k)
&=& 2\cos \frac{k}{2} \sigma_x^\text{eff}(k,d) - 2\delta \sin \frac{k}{2} \sigma_y^\text{eff}(k,d) + \Delta \sigma_z .\nn \\
&&
\end{eqnarray}

Following~\cite{Vanderbilt1993}, a convenient parametrization of the model is in terms of an energy $M>0$ and an angle $\theta$ such that $\Delta=M \cos \theta$ and $\delta=M\sin \theta$. In the following, we fix $M=0.6$ and vary $\theta$. Particular values with inversion symmetry are $\theta=0,\pi$ (CDW) and $\theta=\pi/2,3\pi/2$ (SSH$_{1/2}$). 

The RM model is invariant under the following parameter transformation $(d,\Delta,\delta)\to (1-d,-\Delta,-\delta)$ and relabeling of the two sublattices ($A\leftrightarrow B$). 
On the Bloch Hamiltonian, this symmetry acts as:
\begin{eqnarray}
H_{d,\Delta,\delta}(k) \to \sigma_x H_{1-d,-\Delta,-\delta}(k) \sigma_x =  H_{d,\Delta,\delta}(k)  \, .
\end{eqnarray}
Without loss of generality, one may therefore restrict the study to $0 \leq \theta <\pi$. 

\begin{figure}
\begin{center}
\includegraphics[width=\linewidth]{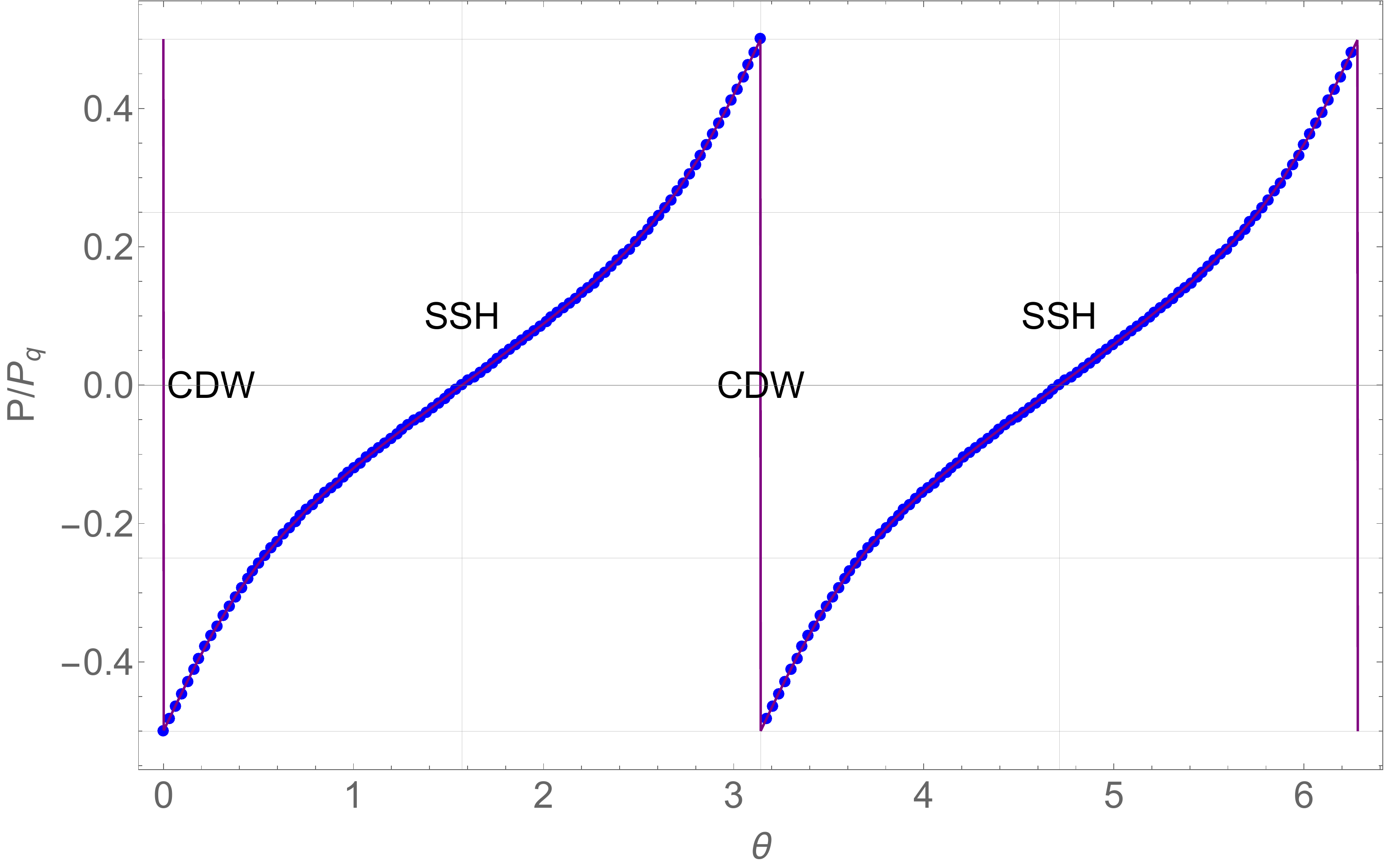}
\caption{\label{fig:bulkpol} Bulk polarization $P$ (in units of $P_q=g/2$) as a function of $\theta$ for the Rice-Mele model with $M=0.6$. (a) Blue dots are for a finite PBC chain of 100 sites. The purple line is Eq.~(\ref{eq:analytic}). 
}
\end{center}
\end{figure}

For the specific ionic model in which every site carries a $g/2$ charge, the bulk polarization is proportional to the modified Zak phase, see Eq.~(\ref{eq:bep}). It was computed numerically in Refs.~\cite{Vanderbilt1993,Xiao2010,Combes2016} for a finite PBC chain with $d=1/2$ assuming that $P_q=g$. It should be modified in order to account for the fact that the quantum of polarization is actually $P_q=g/2$. Here, we also compute it analytically directly in the thermodynamic limit to find
\begin{eqnarray} \label{eq:analytic}
P
&=&-\frac{g}{4}\left[1-\text{sgn }\delta \right.  \\
&+&\left. \frac{\Delta \delta}{\pi\sqrt{1+\Delta^2/4}}\Pi \left(1-\delta^2,\frac{1-\delta^2}{1+\Delta^2/4}\right)\right]  \text{ mod }\frac{g}{2}\, , \nn
\end{eqnarray}
where $\Pi(n,m)$ is the complete elliptic integral of third kind. Numerical and analytical results are plotted in Fig.~\ref{fig:bulkpol}. We see that $P(-\Delta,-\delta)= P(\Delta,\delta)    \text{ mod } P_q$ in agreement with the relabelling symmetry. In particular, for SSH$_{1/2}$ $P=0$ mod $P_q$ (i.e. $\vartheta=0$) and $P=P_q/2$ mod $P_q$ for CDW (i.e. $\vartheta=\pi$). 

As a last confirmation that $P_q=g/2$, consider the polarization computed in~\cite{Vanderbilt1993,Combes2016} for the RM model as a function of $\theta$ at fixed $M$ (see Fig.~4 in~\cite{Vanderbilt1993} and Fig.~3 in~\cite{Combes2016}). In the four cases that have inversion symmetry ($\theta=0,\pi/2,\pi,3\pi/2$), the polarization was found to be $-g/4$, $0$, $g/4$ and $g/2$. But a centro-symmetric crystal must satisfy $P= m P_q/2$, where $m \in \mathbb{Z}$. We therefore have to conclude that $P_q=g/2$ such that $0\equiv g/2$ and $g/4\equiv - g/4$.

\end{appendix}


\begin{thebibliography}{99}

\bibitem{Haldane1988}F. D. M. Haldane, \textit{Model for a Quantum Hall Effect without Landau Levels: Condensed-Matter Realization of the ``Parity Anomaly"}, \href{https://doi.org/10.1103/PhysRevLett.61.2015}{Phys. Rev. Lett. \textbf{61}, 2015 (1988)}.

\bibitem{Fuchs2010}J.-N. Fuchs, F. Pi\'echon, M.O. Goerbig and G. Montambaux, \textit{Topological Berry phase and semiclassical quantization of cyclotron orbits for two dimensional electrons in coupled band models}, \href{https://doi.org/10.1140/epjb/e2010-00259-2}{Eur. Phys. B \textbf{77}, 351 (2010)}.

\bibitem{Fruchart2014}M. Fruchart, D. Carpentier and K. Gawedzki, \textit{Parallel Transport and Band Theory in Crystals}, \href{https://doi.org/10.1209/0295-5075/106/60002}{Europhys. Lett. \textbf{106}, 60002 (2014)}.

\bibitem{Lim2015}L.-K. Lim, J.-N. Fuchs and G. Montambaux, \textit{Geometry of Bloch states probed by St\"uckelberg interferometry}, \href{https://doi.org/10.1103/PhysRevA.92.063627}{Phys. Rev. A \textbf{92}, 063627 (2015)}.

\bibitem{Simon2020}S. H. Simon and M. Rudner, \textit{Contrasting lattice geometry dependent versus independent quantities: Ramifications for Berry curvature, energy gaps, and dynamics}, \href{https://doi.org/10.1103/PhysRevB.102.165148}{Phys. Rev. B \textbf{102}, 165148 (2020)}.

\bibitem{Cayssol2021}J. Cayssol and J.-N. Fuchs, \textit{Topological and geometrical aspects of band theory}, \href{https://doi.org/10.1088/2515-7639/abf0b5}{J. Phys. Mater. \textbf{4}, 034007 (2021)}.

\bibitem{Schnyder2008}A. P. Schnyder, S. Ryu, A. Furusaki and A. W. W. Ludwig, \textit{Classification of topological insulators and superconductors in three spatial dimensions}, \href{https://doi.org/10.1103/PhysRevB.78.195125}{Phys. Rev. B \textbf{78}, 195125 (2008)}.

\bibitem{Kitaev2009}A. Kitaev, \textit{Periodic table for topological insulators and superconductors}, \href{https://doi.org/10.1063/1.3149495}{AIP Conference Proceedings \textbf{1134}, 22 (2009)}.

\bibitem{Chiu2016}C.-K. Chiu, J. C. Y. Teo, A. P. Schnyder, and S. Ryu, \textit{Classification of topological quantum matter with symmetries}, \href{https://doi.org/10.1103/RevModPhys.88.035005}{Rev. Mod. Phys. \textbf{88}, 035005 (2016)}.

\bibitem{Bena2009}C. Bena and G. Montambaux, \textit{Remarks on the tight-binding model of graphene}, \href{https://doi.org/10.1088/1367-2630/11/9/095003}{New J. Phys. \textbf{11}, 095003 (2009)}.

\bibitem{Vanderbilt2018}D. Vanderbilt, \textit{Berry phase in electronic structure theory: electric polarization, orbital magnetization and topological insulators} \href{https://doi.org/10.1017/9781316662205}{(Cambridge university press, 2018)}.

\bibitem{SSH}W. P. Su, J. R. Schrieffer, and A. J. Heeger, \textit{Solitons in Polyacetylene}, \href{https://doi.org/10.1103/PhysRevLett.42.1698}{Phys. Rev. Lett. \textbf{42}, 1698 (1979)}.

\bibitem{Hughes2011}T. L. Hughes, E. Prodan and B. A. Bernevig, \textit{Inversion-symmetric topological insulators}, \href{https://doi.org/10.1103/PhysRevB.83.245132}{Phys. Rev. B \textbf{83}, 245132 (2011)}.

\bibitem{ChenLee2011}K. T. Chen and P. A. Lee, \textit{Static electric field in one-dimensional insulators without boundaries}, \href{https://doi.org/10.1103/PhysRevB.84.113111}{Phys. Rev. B {\bf 84}, 113111 (2011)}.

\bibitem{Zak1989}J. Zak, \textit{Berry's phase for energy bands in solids}, \href{https://doi.org/10.1103/PhysRevLett.62.2747}{Phys. Rev. Lett. \textbf{62}, 2747 (1989)}.

\bibitem{Zak1989b}J. Zak, \textit{Berry's Geometrical Phase for Noncyclic Hamiltonians}, \href{https://doi.org/10.1209/0295-5075/9/7/001}{Europhys. Lett. \textbf{9}, 615 (1989)}.

\bibitem{Shockley}W. Shockley, \textit{On the Surface States Associated with a Periodic Potential}, \href{https://doi.org/10.1103/PhysRev.56.317}{Phys. Rev. \textbf{56}, 317 (1939)}.

\bibitem{Vanderbilt1993}D. Vanderbilt and R. D. King-Smith, \textit{Electric polarization as a bulk quantity and its relation to surface charge}, \href{https://doi.org/10.1103/PhysRevB.48.4442}{Phys. Rev. B \textbf{48}, 4442 (1993)}.

\bibitem{RiceMele1982}M. J. Rice and E. J. Mele, \textit{Elementary Excitations of a Linearly Conjugated Diatomic Polymer}, \href{https://doi.org/10.1103/PhysRevLett.49.1455}{Phys. Rev. Lett. \textbf{49}, 1455 (1982)}.

\bibitem{Resta2000}R. Resta, \textit{Manifestations of Berry's phase in molecules and condensed matter}, \href{https://doi.org/10.1088/0953-8984/12/9/201}{J. Phys.: Condens. Matter \textbf{12}, R107 (2000)}.

\bibitem{footnotegauge}In the periodic gauge $|u_n(k+G)\rangle=e^{-i G x}|u_n(k)\rangle$~\cite{Resta2000}, where $G$ is a reciprocal lattice vector, the closure term $\langle u_n(-\pi)|e^{i 2\pi x} |u_n(\pi)\rangle=1$ and one recovers the standard expression for the Zak phase $Z_n= \int_{-\pi}^\pi  dk \langle u_n |i \partial_k u_n\rangle$.

\bibitem{King-Smith1993}R. D. King-Smith and D. Vanderbilt, \textit{Theory of polarization of crystalline solids}, \href{https://doi.org/10.1103/PhysRevB.47.1651}{Phys. Rev. B \textbf{47}, 1651(R)  (1993)}.

\bibitem{Resta2007}R. Resta and D. Vanderbilt, \textit{Theory of Polarization: A Modern Approach}, in Physics of Ferroelectrics: A Modern Perspective, Topics in Applied Physics Vol. 105 (Springer-Verlag, Berlin, Heidelberg, 2007), pp. 21-68.

\bibitem{footnoteposition}The dependence of the modified Zak phase $\bar{Z}=Z-2\pi \bar{x}$ on the position operator $x$ is not only through $\bar{x}$, but also through the cell periodic Bloch states used in the computation of the Zak phase. They are related to the Bloch states $|\psi_{nk}\rangle$, which only depend on $H$, by $|u_n(k)\rangle = e^{-i k x} |\psi_{nk}\rangle$.

\bibitem{footnoteminus}The minus sign in the definition of $\vartheta$ is to make a link with the polarization angle usually defined in the effective electromagnetic Lagrangian of 1D dielectrics, see Appendix~\ref{sec:ep}.

\bibitem{footnotez2}This $\mathbb{Z}_2$ invariant is related to inversion symmetry and should be clearly distinguished from the spontaneous $\mathbb{Z}_2$ symmetry breaking occurring at the Peierls transition.

\bibitem{Qi2008}X.-L. Qi, T. L.Hughes and S.-C. Zhang, \textit{Topological field theory of time-reversal invariant insulators}, \href{https://doi.org/10.1103/PhysRevB.78.195424}{Phys. Rev. B \textbf{78}, 195424 (2008)}.

\bibitem{Cooper2019}N. R. Cooper, J. Dalibard and I. B. Spielman, \textit{Topological bands for ultracold atoms}, \href{https://doi.org/10.1103/RevModPhys.91.015005}{Rev. Mod. Phys. \textbf{91}, 015005 (2019)}.

\bibitem{Yakovenko2012}S. S. Pershoguba and V. M. Yakovenko, \textit{Shockley model description of surface states in topological insulators}, \href{https://doi.org/10.1103/PhysRevB.86.075304}{Phys. Rev. B \textbf{86}, 075304 (2012)}.

\bibitem{Bradlyn2017}B. Bradlyn, L. Elcoro, J. Cano, M. G. Vergniory, Z. Wang, C. Felser, M. I. Aroyo and B. A. Bernevig, \textit{Topological quantum chemistry}, \href{https://doi.org/10.1038/nature23268}{Nature \textbf{547}, 298 (2017)}.

\bibitem{footnotezakwinding}In many references, what is called a Zak phase is actually $\pi$ times a winding number defined with respect to a specific unit cell choice imposed by an edge. Its computation is usually done with a periodic (rather than canonical) Bloch Hamiltonian. It is not the phase $Z_n$ defined by Zak~\cite{Zak1989, Zak1989b}, and used here, which requires us to use the canonical Bloch Hamiltonian such as to have the proper relation between the Zak phase and the Wannier center. In particular, the Zak phase depends continuously on the position origin, which would be absurd for a winding number.

\bibitem{Creutz1999}M. Creutz, \textit{End States, Ladder Compounds, and Domain-Wall Fermions}, \href{https://doi.org/10.1103/PhysRevLett.83.2636}{Phys. Rev. Lett. \textbf{83}, 2636 (1999)}.

\bibitem{Kitaev2001}A. Yu. Kitaev, \textit{Unpaired Majorana fermions in quantum wires}, \href{https://doi.org/10.1070/1063-7869/44/10S/S29}{Phys.-Usp. \textbf{44}, 131 (2001)}.

\bibitem{Guzman2020}M. Guzm\'an, D. Bartolo and D. Carpentier, \textit{Geometry and topology tango in chiral materials}, \href{https://arxiv.org/abs/2002.02850}{arXiv:2002.02850} to appear in SciPost).

\bibitem{Jackiw1976}R. Jackiw and C. Rebbi, \textit{Solitons with fermion number 1/2}, \href{https://doi.org/10.1103/PhysRevD.13.3398}{Phys. Rev. D \textbf{13}, 3398 (1976)}.

\bibitem{Ryu2002}S. Ryu and Y. Hatsugai, \textit{Topological Origin of Zero-Energy Edge States in Particle-Hole Symmetric Systems}, \href{https://doi.org/10.1103/PhysRevLett.89.077002}{Phys. Rev. Lett. \textbf{89}, 077002 (2002)}.

\bibitem{Delplace2011}P. Delplace, D. Ullmo, and G. Montambaux, \textit{Zak phase and the existence of edge states in graphene}, \href{https://doi.org/10.1103/PhysRevB.84.195452}{Phys. Rev. B \textbf{84}, 195452 (2011)}.

\bibitem{footnotecurvature}The Chern number (topological invariant) does not depend on the spatial embedding of the orbitals. However, the Berry curvature (geometrical quantity) does depend on it, see~\cite{Fuchs2010, Fruchart2014,Lim2015}.

\bibitem{Atala2013}M. Atala, M. Aidelsburger, J. T. Barreiro, D. Abanin, T. Kitagawa, E. Demler and I. Bloch, \textit{Direct measurement of the Zak phase in topological Bloch bands}, \href{https://doi.org/10.1038/nphys2790}{Nature Phys. \textbf{9}, 795 (2013)}.

\bibitem{Dutreix2020}C. Dutreix, M. Bellec, P. Delplace and F. Mortessagne, \textit{Wavefront dislocations reveal insulator topology}, \href{https://doi.org/10.1038/s41467-021-23790-w}{Nature Comm. \textbf{12}, 3571 (2021)}.

\bibitem{StJean2017} P. St-Jean, V. Goblot, E. Galopin, A. Lema\^itre, T. Ozawa, L. Le Gratiet, I. Sagnes, J. Bloch and A. Amo, \textit{Lasing in topological edge states of a one-dimensional lattice}, \href{https://doi.org/10.1038/s41566-017-0006-2}{Nat. Photon. \textbf{11}, 651 (2017)}.

\bibitem{Kudin2007}K. N. Kudin, R. Car and R. Resta, \textit{Quantization of the dipole moment and of the end charges in push-pull polymers}, \href{https://doi.org/10.1063/1.2799514}{J. Chem. Phys. \textbf{127}, 194902 (2007)}.

\bibitem{Xiao2010}D. Xiao, M. C. Chang and Q. Niu, \textit{Berry phase effects on electronic properties}, \href{https://doi.org/10.1103/RevModPhys.82.1959}{Rev. Mod. Phys. \textbf{82}, 1959 (2010)}.

\bibitem{Combes2016} F. Combes, M. Trescher, F. Pi\'echon and J.-N. Fuchs, \textit{Statistical mechanics approach to the electric polarization and dielectric constant of band insulators}, 
 \href{https://doi.org/10.1103/PhysRevB.94.155109}{Phys. Rev. B \textbf{94}, 155109 (2016)}.




\end{thebibliography}
\end{document}